\documentclass[twocolumn,letterpaper]{IEEEAerospaceCLS}  

\usepackage[]{graphicx}    
\usepackage{subfig}
\usepackage{amsmath}
\usepackage{verbatim}
\usepackage{graphicx}
\usepackage{float}
\usepackage[utf8]{inputenc}
\usepackage{color}
\newcommand{\ignore}[1]{}  

\begin{document}
\title{Energy Efficient Routing For Underwater Acoustic Sensor Network Using Genetic Algorithm}

\author{%
Arjun Prasad Chaurasiya\\ 
Department of Computer Science \& Engineering\\
Vel Tech University\\
Chennai, India \\
chaurasiyaarjun7@gmail.com
\and 
Roshan Sah\\ 
Department of Aerospace Engineering\\
Indian Institute of Technology, Kharagpur\\
Kharagpur, India \\
sahroshan11@gmail.com
\and
Dr.V.Sivakumar\\ 
Department of Computer Science \& Engineering\\
Vel Tech University\\
Chennai, India \\
drsivakumarv@veltech.edu.in
}

\maketitle

\thispagestyle{plain}
\pagestyle{plain}

\maketitle

\thispagestyle{plain}
\pagestyle{plain}

\begin{abstract}
In underwater acoustic sensor networks (UWASN), energy-reliable data transmission is a challenging task . This is due to acoustic transmission disturbances caused by excessive noise, exceptionally long propagation delays, a high bit error rate, limited bandwidth capability, and interference.  One of the most important issues of UWASN for research is how to extend life span of data transmission. Data transfer from a source node to a destination node in UWASN is a complicated topic for researchers. Many routing algorithms, such as vector base forwarding and depth base routing, have been developed in past years.We propose a genetic algorithm-based optimization method for improving the energy efficiency of data transmission in the routing Path from a source node to a destination node.

\end{abstract}

\tableofcontents

\section{Introduction}
\label{Introduction}
According to a new survey undertaken by the United States National Oceanic and Atmospheric Administration (NOAA), oceans occupy nearly 97 percent of the Earth's surface \cite{noaa}. There is an increasing interest in studying the underwater environment for a variety of purposes, including climate change, oceanic animal research, oil platform tracking, surveillance, and unmanned operations. All of these applications involve a medium of communication both inside the underwater environment and the rest of the world. As a result, research into underwater wireless communication has gained a lot of interest.

Communication devices based on acoustic waves, high frequency waves, and optical waves are used to introduce underwater wireless communications today. Since they can communicate over long distances, underwater acoustic wireless networks have become one of the most commonly used underwater wireless communications systems. In UWASN which are made up of several autonomous and self-organizing sensor nodes. To collect exact information from deep water, these nodes are manually dispersed in various depths in underwater environments. Collecting data to a sink on the water's surface through acoustic waves. The normal sensor nodes are present in these networks, along with an acoustic modem for connectivity. Sinks are fitted with both acoustic and radio modems, allowing them to receive data from underwater nodes via acoustic waves and transmit it via radio waves to the base station.

The traditional method of disaster control is to deploy an underwater sensor and manually collect data \cite{https://doi.org/10.1002/dac.3445}. The disadvantage of the traditional approach of not being able to provide timely information inspired a wave of interest in the concept of an underwater sensor network. To deal with these disasters using an underwater sensor network, a reliable communication medium to exchange data is needed. Electromagnetic waves are initially effective for short-distance communication, but due to high attenuation in the sea, they are insufficient for long-distance communications \cite{https://doi.org/10.1002/dac.3303}, while optical waves are easily absorbed and dispersed in the water, allowing for very short-distance communication \cite{RANI201742}. Acoustic waves resolve the above limitations and can transmit over long distances with low frequency absorption and attenuation \cite{8675480}. Since an acoustic wave propagates as a pressure wave, it may transmit its energy over long distances. However, because of the low speed of the audio signal underwater (approximately 1500 m/s), the acoustic signal has its own drawbacks, such as restricted bandwidth [BW] and long transmission delay \cite{s16010098}.

Several quality of service awarerouting protocols, such as vector-based forwarding (VBF) \cite{10.1007/11753810_111} and depth-based forwarding (DBR) \cite{10.1007/978-3-540-79549-0_7}, have been proposed for UWASN in recent years. In terrestrial wireless sensor networks, a wide number of network applications have been proposed for determining a route from source to sink.These protocols are built on an end-to-end approach, which is ineffective in high-dynamic topology networks with long propagation delays. Since UWASN is such a new topic in this field, most researchers are focusing on the physical layer, connection layer, and localization, while network layer analysis is still in its infancy. As a result, only a few routing protocols for UWASN have been established.
Optimization technique to increase the energy efficiency of data transmission in routing Path from a source node to a destination node in UWASN is the most promising approach in an underwater environment.

 This paper is organized as follows: Section I presents the Introduction model, Section II present the Literature Review, Section III presents the Proposed Routing Scheme, Section IV present Implementation model, Section V present the performance evaluation study. Finally, Section VI Conclusion and Future Enhancements.

\vspace{-5mm}
\section{Literature Review}
\label{LITERATURE REVIEW}
In the field of underwater acoustic sensor network, several studies have used a number of ideas and methods that have proven to be successful. Many researchers have proposed a number of  algorithms in underwater acoustic sensor network ; we've briefly listed a few of them below.

A. Wahid et al.\cite{6983274} propose a functional node cooperation (NC) protocol to improve collection efficiency by taking advantage of the fact that underwater nodes can hear other nodes' transmissions.The underwater data collection area is divided into several sub-zones to minimise the source level of underwater nodes, but in each sub-zone, the mobile surface node using the NC protocol could move adaptively between selective relay cooperation (SRC) and dynamic network coded cooperation (DNC).

X. Zhuo et al. \cite{8698756} purpose of this report is on the use of underwater acoustic sensor networks (UASNs), which are distinguished by their large scale, dispersed distribution, and differing traffic loads. The strategy to access the popular communication medium is needed to improve the efficiency of UASNs because the underwater sensor channel is known for its restricted bandwidth, time variation, and high propagation delays.

M. Molins et al. \cite{4393832} proposed a new protocol is known as slotted FAMA because it employs time slotting. Time slotting reduces the need for unnecessarily long control packets, resulting in energy savings. Via simulation of a mobile ad hoc underwater network, protocol efficiency in terms of throughput and delay is evaluated, demonstrating the presence of an acceptable rated power to be used for a given user density.

Hongyu Cui et al.\cite{7535751} present the platforms built by our lab, which are based on the well-known ns2 and ns-miracle simulators and feature a layered structure that supports cross-layer signalling. The simulation platform runs on a machine with the Linux operating system installed. The platform's architecture is divided into four layers: an application layer, a routing layer, a medium access layer, and a data access layer.

Jun-Hong et al.\cite{1637927} present a novel networking model for exploring aqueous environments is the large-scale mobile underwater wireless sensor network (UWSN). Remote UWSNs, on the other hand, vary greatly from ground-based wireless sensor networks in terms of communication bandwidth, propagation delay, moving node mobility, and failure probability.

Z. Wu et al.\cite{7084366} presented a new underwater acoustic network localization algorithm is implemented. To improve the sensitivity of random noise, the algorithm uses the raw data before calculating the localization. We use a more reliable set of coordinates and a modified calculation technique to reduce the consistency of the calculation results. The new algorithm is better suited to the underwater acoustic sensor network.

Present Aqua-Sim,P. Xie et al.\cite{5422081} a network simulator for underwater sensor networks. Aqua-Sim is based on NS-2, among the most commonly used network simulators. It is constructed in an object-oriented manner, with all network entities configured as classes. Aqua-Sim accurately simulates underwater acoustic channel attenuation and impact behaviours in long delay acoustic networks.

Our method is implemented T. Hu et al.\cite{4745119} on standard MAC protocols and aims to increase the lifespan of networks by uniformly distributing residual energy from sensor nodes. All throughout routing process, the energy consumption of each node, and also the power generation among a group, is taken into account to measure the reward function, which aids in the selection of appropriate packet forwarders.

The Yi Cui et al.\cite{1634912} maximum lifetime routing problem is given a novel utility-based nonlinear formulation. A completely distributed localised routing algorithm based on this concept is also introduced, and it is shown to converge at the optimum stage, where the network lifetime is maximised. To validate the proposed solution, proper mathematical analysis and simulator results are presented.

A. Sankar et al.\cite{1356995} propose a distributed routing scheme that, in a linearly low percent error, reaches the optimal solution. Our method is formulated on the basis of a multi-commodity flow, which allows various power consumption models and bandwidth constraints to be considered. It works with both static and dynamic networks that shift slowly.

V. D. Park et al.\cite{631180} presented for remote, multihop wireless networks, we present a new distributed routing mechanism. The protocol is part of a group of algorithms known as "connection reversal" algorithms. The protocol's response is organised as a time-ordered series of dispersing calculations, each of which is made up of a set of guided connection reversals.

\vspace{-4mm}
\section{Proposed Routing Scheme}
\label{Proposed Routing Scheme}

To achieve cooperative sampling of a two-dimensional (2-D) ocean scenario, a network of underwater sensors at various depths is deployed. The following properties are considered in this model. First, in term of transmission range, all randomly deployed sensor nodes have the same capabilities and located near sea surface  can communicate with each other directly.Second, each sensor node uses a location information to determine its own location. Third, each node not only travels in a specific horizontal position, but also slightly in a vertical position. Fourth, the acoustic channel is symmetric, the energy needed to transmit a message from source to destination is similar.Fifth, each sensor node will change its low and high transmission power periodically. Furthermore, it is believed that the sink node has a lot of resources and can manage many packets at about the same time.

\subsection{A. Modifed Genetic Algorithm}
\label{Modifed Genetic Algorithm}

The Genetic Algorithm (GA) is a type of evolutionary heuristic search algorithm that belongs to the evolutionary algorithms group. The Genetic Algorithm is a search-driven optimization algorithm focused on genetics and natural selection concepts. It's often used to find ideal or near-optimal solutions to complex problems that would require a lifetime to solve. It's widely used to solve optimization problems, as well as in science and machine learning. They're widely used to come up with high-quality solutions to optimization and search issues. Genetic Algorithms are capable of delivering a "good-enough" solution "quickly enough." As a consequence, genetic algorithms are appealing for solving complex optimization problems.

\begin{figure}
    \centering
    \includegraphics[width=0.75\linewidth]{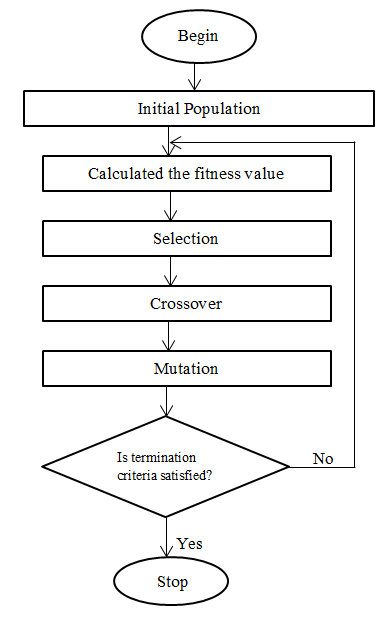}
    \caption{Genetic Algorithm}
    \label{problem}
\end{figure}

\subsection{Flowchart : }
\label{Modifed Genetic Algorithm}

\subsubsection{Begin}
A population of n chromosomes is generated at random.

\subsubsection{Initial population} 
The phase starts with a group of people known as a Population. Each person is a possible solution to the problem you're trying to solve. Genes are a collection of parameters that define an organism.

\subsubsection{Fitness} 
Chromosome's fitness is calculated by an assigned fitness work. The fitness function defines a person's level of fitness. It assigns each individual a fitness score. The fitness score determines the chances of a person being chosen for reproduction.

\subsubsection{Selection} 
Choose two chromosomes with the highest wellness prediction from the population. The aim of the selection process is to find the fittest individuals and allow them to transfer their genes down to future generations.

\subsubsection{Crossover}  
A genetic algorithm's most important step is crossover. A crossover point is selected at random from within the genes for each pair of parents to be paired up. Parents' genes are shared among themselves to produce offspring until the convergence point is reached. For example –

\begin{figure}
    \centering
    \includegraphics[width=0.75\linewidth]{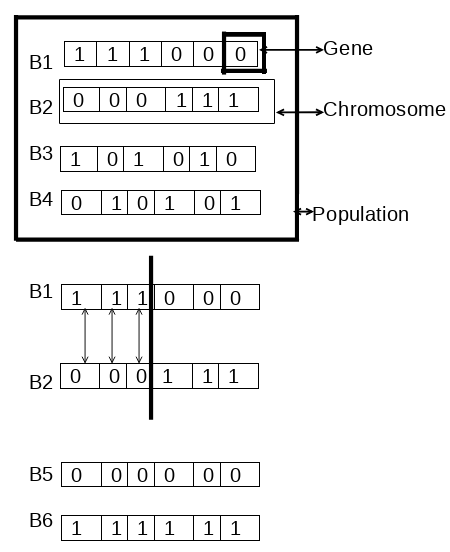}
    \caption{Genetic Algorithm}
    \label{problem}
\end{figure}

\subsubsection{Selection} 
Any of the genes of certain new offspring may be exposed to a mutation with a low crossover sampling. This means that some of the bits in the bit string can be switched around. A mutation changes the value of one or more genes in a chromosome from its original state. Mutation occurs to maintain diversity of the population.

\subsubsection{Termination} 
The population is reintroduced with the latest offspring. If the population has converged, the algorithm will end. The genetic algorithm is being said to have given a series of approaches to our problem.

\subsection{Difference between Underwater Acoustic Sensor Network(UWASN) and  Terrestrial Wireless Sensor Network(TWSN) }

 In terms of various factors such as environmental conditions and communication medium, UWASN differs significantly from TWSN, resulting in unique characteristics and challenges. There are a few main differences are:

\begin{table}[htbp]
\caption{Comparison of UWASN and TWSN }

\begin{center}
\begin{tabular}{|c|c|c|c|}
\hline
\textbf{S.No.} & \textbf{\textbf{Feature }}& \textbf{\textbf{UWASN}}& \textbf{\textbf{TWSN}}\\
\hline
1 & Architecture & Mostely 3D & Mostely 2D   \\
\hline
2 & Node movement & Mobility & Fixed   \\
\hline
3 & Propagation delay & high & low \\
\hline
4 & Frequency & Low & High\\
\hline
5 & Topology &  High dynamic & Low \\
\hline
6 & Communication &  Acoustic & Radio\\
\hline
7 & Path loss & High & Low\\
\hline
8 & Price & Expensive & Cheap\\
\hline
9 & Range & Vast Area & Small Area\\
\hline
10 & Sensor size & Large & Small \\
\hline
\end{tabular}
\label{tab1}
\end{center}
\end{table}

\section{Implementation}
\label{IMPLEMENTATION}
The execution of these modules takes place on a terminal in a VMware sensor machine, with input formats in the Network Simulator Version tcl format.

\begin{figure}
    \centering
    \includegraphics[width=0.95\linewidth]{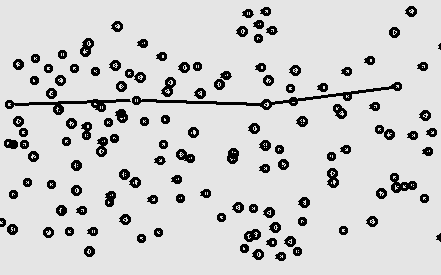}
    \caption{The nodes is moving towards the destination.}
    \label{IMPLEMENTATION}
\end{figure}

\subsection{Module 1: Configure the data and run in the terminal}
The data is installed and stored in the library feature in this module, which can be used for future reference. The data code is configured using ns2 scripting, which is run in the terminal as "$\#$configure." It goes through all of the data that needs to be modified and then moves on to the next version after all of the data has been configured.
\begin{figure}
    \centering
    \includegraphics[width=0.95\linewidth]{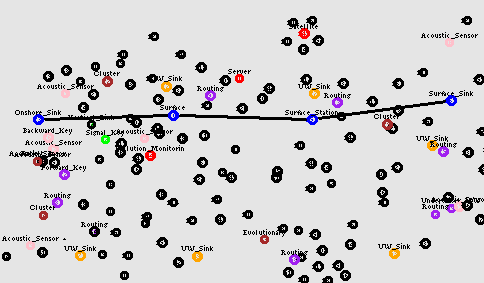}
    \caption{The nodes is moving towards the destination.}
    \label{IMPLEMENTATION}
\end{figure}

\subsection{Module 2: Launch  the “nan” in the terminal }
After configuring the data in the terminal, which tests and collects all of the data, the process continues by installing nan in the terminal, which includes all of the physical data constraints as well as a graphical representation of the node mobility from source to destination.
\vspace{-3mm}
\begin{figure}
    \centering
    \includegraphics[width=0.95\linewidth]{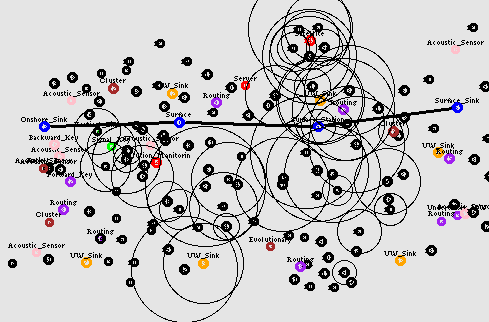}
    \caption{Node moves from source node to  destination node.}
    \label{IMPLEMENTATION}
\end{figure}

\subsection{Module 3: Final stage of simulation }
By running “ns project.tcl” in the terminal, which includes the entire simulation module, the ended stage of the simulation component has been achieved in this phase. This model includes all of the physical sorting lists that must be built in order for the node movement to operate in a network animator form, which is then visually presented.

\subsection{Module 4: Node movement in nam}
The node animator is used in the final phase of the system to display node movement as it travels from source to destination through an underwater acoustic sensor network. In the terminal, type "nam out.nam" to see the network animator's dialogue while watching the node movement.
When the simulation is turned on, the nodes can travel from the source to the destination in a specific amount of time in order to save energy for later use. When the node arrives at its destination, the various movements of the node are recorded in a visual analysis.

\section{Performance Evaluation}
\label{Performance}

PDR in all routing schemes gradually increases with the rise in round numbers between 1 and 100, as shown in fig.6. When the network size is low, between 1 and 35, the VBF routing scheme performs better than the DBR routing scheme in order to achieve higher PDR.
When the size of the network is between 36 and 70, DBR performs slightly better than VBF in terms of achieving a higher PDR. When the network size becomes large between round numbers 71 and 100.
The PDR of EER is found to be better than DBR and VBF routing schemes in all regions. On the other hand, we find that DBR has a better PDR than VBF because data packets are forwarding over better links among nodes in UWASN.

The efficiency of all routing schemes rapidly decreases with the increase in nodes between 1 and 76, as shown in Fig.7. When the network has up to 76 sensor nodes, the delays rate efficiency of DBR is better than VBF.In all three different regions, EER outperforms both DBR and VBF routing schemes in terms of delay efficiency. We also note that DBR has a better delay performance than VBF because it considers the shortest path when forwarding data packets.

As shown in Fig. 8, the residual energy of sensor nodes decreases as the round number between 1 and 100 increases in all of the routing schemes. Although VBF's energy consumption profile is better than DBR's, DBR's residual energy output is better in round number between 36 and 70 and between 71 and 100.
In terms of residual energy efficiency, however, EER performs better both DBR and VBF because it keeps sensor nodes alive throughout the rounds.

\begin{figure}
    \centering
    \includegraphics[width=0.95\linewidth,height =2.5 in]{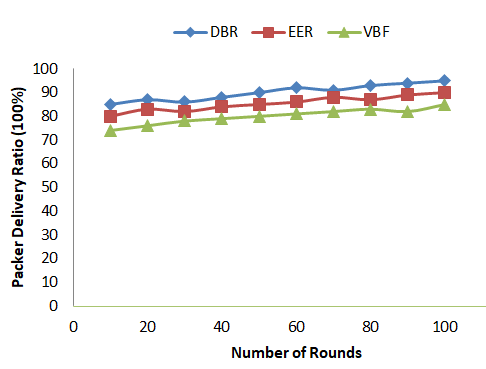}
    \caption{PDR versus round numbers between 1 and 100.}
    \label{IMPLEMENTATION}
\end{figure}

 Finally, due to its novel genetic algorithm created to organise sensor nodes into a connected structure in order to balance the load and prolong the network lifetime. EER performs better than both DBR and VBF in terms of network lifespan, as shown in Fig.9.

In Underwater Networking, as shown in Fig.10, Lifetime excellent efficiency in terms of packet delivery ratio is achieved. For real-time UWASN-based underwater applications, we have successfully enhanced data transfer reliability for energy-delay. In the highly dynamic underwater environment, EER significantly decreases the possibility of packet loss and maintains high connection quality among nodes.

\begin{figure}
    \centering
    \includegraphics[width=0.95\linewidth, height =2.5 in]{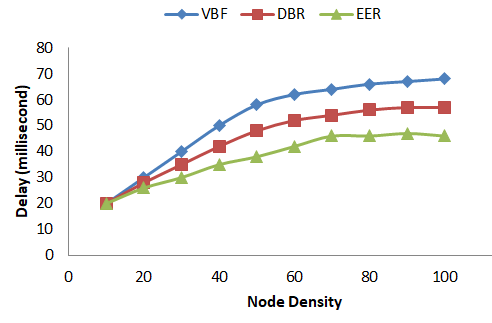}
    \caption{The delay versus node density between 1 and 76.}
    \label{IMPLEMENTATION}
\end{figure}

\begin{figure}
    \centering
    \includegraphics[width=0.95\linewidth,height =2.5 in]{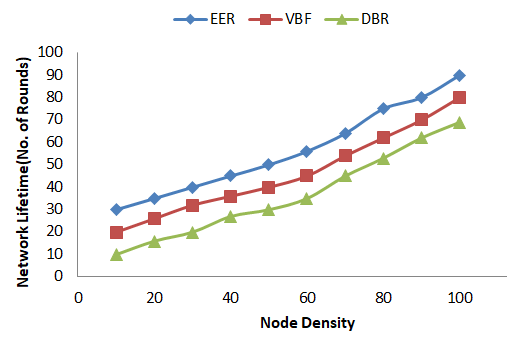}
    \caption{The residual energy versus number of round numbers between 1 and 100.}
    \label{IMPLEMENTATION}
\end{figure}

\begin{figure}
    \centering
    \includegraphics[width=0.95\linewidth, height =2.5 in]{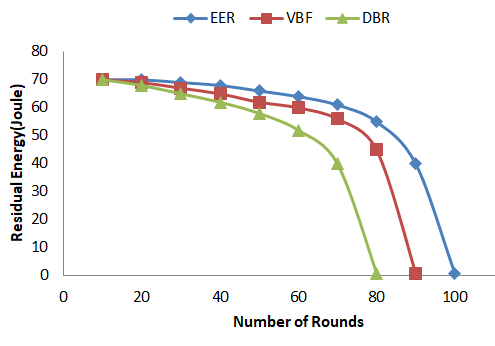}
    \caption{Network lifetime versus node density between 1 and 76. }
    \label{IMPLEMENTATION}
\end{figure}

\begin{figure}
    \centering
    \includegraphics[width=0.95\linewidth,height =2.5 in]{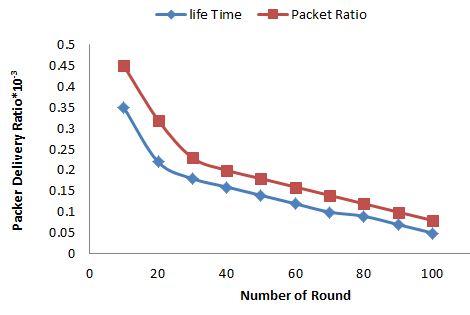}
    \caption{ Underwater Networking.}
    \label{IMPLEMENTATION}
\end{figure}

\section{Conclusion}
\label{Conclusion}

In this paper we have described genetic algorithm to transfer a data from source node to destination node for  underwater acoustic sensor network applications is improved. Due to the growing propagation delays in the underwater environment and the medium's characteristic of being unable to detect any collision, when any data is transmitted using the wireless medium, energy consumption increases and data throughput falls.
The results of the performance show that EER performs exceptionally well in terms of metrics such as PDR, average end-to-end packet loss, and energy consumption.

\section{Future Enhancements}
\label{Conclusion}

Future research will focus on increasing the quality and reliability of underwater communication networks. The bandwidth, connection quality, and bit error rate in UWASN are all lower than those of terrestrial radio frequency channels, and work needs to be done to improve them.

\bibliographystyle{ieeetr}
\bibliography{sample}



\end{document}